\documentclass {article}

\usepackage {amsmath}
\usepackage {amssymb}
\usepackage {amsthm}
\usepackage {amsfonts}
\usepackage {ascmac}
\usepackage {ulem}
\usepackage [bbgreekl]{mathbbol}
\usepackage [dvipdfmx]{graphicx}
\usepackage {caption}

\newtheorem{Def}{\underline{Definition}}
\newtheorem{Lem}{\underline{Lemma}}

\begin{document}

\title{Universality for random tensors and \\cycle graphs with multiple edges}
\date{\empty}
\author{Nana Kanbe}
\maketitle
\begin{center}\it Graduate School of Mathematics, Nagoya University,
Chikusa-ku,\\ Nagoya 464-8602, Japan \end{center}
\begin{abstract}
We consider the universality for the trace invariants of 
$c_1 N \times \cdots \times c_D N$ tensors with i.i.d. complex random 
 elements. In the case $c_1 = \cdots = c_D$, Gurau derived the 
 universality in the limit $N \to \infty$ by representing 
 the average trace invariants in terms of the corresponding colored graphs. 
 Moreover he could explicitly calculate the asymptotic forms 
 of the average trace invariants, if the 
 corresponding graphs were special ones called "melonic graphs". 
 In this paper we study another kind of special graphs, 
 cycle graphs with multiple edges. 
 One can construct these graphs by 
 using simple cycle graphs and melonic graphs. 
 Counting the number of the components, 
 we can explicitly calculate the asymptotic forms 
 of the average trace invariants.
\end{abstract}

\newpage
Let us suppose that a $c_1N \times \cdots \times c_DN$ tensor $T$ 
has random complex elements ($N$ and $c_i N$ are positive integers).
The real and imaginary parts of the tensor elements are i.i.d.
random numbers with zero odd moments and finite even moments. 
The trace invariants of such a tensor are defined as
 \begin{equation}
 Tr(T,\bar{T})=
 \sum_{n^1,\bar{n}^1=1}^{c_1N}
 \cdots
 \sum_{n^D,\bar{n}^D=1}^{c_DN}
 \prod_{j=1}^k
  \delta_{n^1_j \bar{n}^1_{\sigma_1 (j)}}
  \cdots
  \delta_{n^D_j \bar{n}^D_{\sigma_D (j)}}
  T_{n^1_j \cdots n^D_j}
  \bar{T}_{\bar{n}^1_j \cdots \bar{n}^D_j},
 \end{equation}
where $\bar{T}$ is the complex conjugate of $T$ 
and $\sigma_1,...,\sigma_D$ are cyclic permutations of length $k$ 
on $\{ 1,...,k \}$. 
A trace invariant is a real number which is invariant 
under the unitary transformations of $T$ . 
We can represent a trace invariant 
as a connected $D$-colored graph defined below. 
\begin{Def}\rm
A graph $B$ satisfying the following conditions is 
a $D$-colored graph \cite{gurau1}. 
Here $V(B)$ and $E(B)$  are the sets 
of the vertices and edges of $B$, respectively.
 \begin{itemize}
 \item All edges $e$ of $B$ have vertices $v$ and $\bar{v}$ at their ends. 
 There exists a partition of $V(B)$, $V(B) = V_1(B) \sqcup V_2(B)$, 
 such that $v \in V_1(B)$ and $\bar{v} \in V_2(B)$ for all $e$. 
 We depict $v$ (white vertices) as white circles and $\bar{v}$ 
 (black vertices) as black circles in figures.
 \item  There exists a partition of $E(B)$, $E(B)=\bigsqcup_{i=1}^D E^i(B)$. 
 The element of $E^i(B)$ are called edges with color $i$.
 \item  Each vertex is at the end of $D$ edges, 
 which have different colors from each other. 
 \end{itemize}
\end{Def}
\noindent
A $D$-colored graph $B$ has $k$ white vertices $v_1,...,v_k$, 
and $k$ black vertices $\bar{v}_{\sigma_i(1)},...,\bar{v}_{\sigma_i(k)}$ 
for each color $i$. 
Let us suppose in the trace invariant that 
$T_{n^1_j...n^D_j}$ and 
$\bar{T}_{\bar{n}^1_{\sigma_1(j)} \cdots \bar{n}^D_{\sigma_D(j)}}$
have contraction with respect to $n^i_j$ and $\bar{n}^i_{\sigma_i(j)}$. 
Then $v_j$ and $\bar{v}_{\sigma_i(j)}$ are directly connected 
by an edge with color $i$ on $B$.
 
By using colored graphs, Gurau showed in a special case 
$c_1= \cdots = c_D =1$ that the average value 
$\mu(Tr(T,\bar{T}))$ of the trace invariants have 
universal asymptotic forms in the limit $N \to \infty$. 
Let us first consider that special case.
In order to calculate $\mu(Tr(T,\bar{T}))$, the cumulants
\begin{equation}
\begin{split}
 &\kappa_{2k}\left(T_{n^1_1 \cdots n^D_1}, \bar{T}_{\bar{n}^1_1 \cdots \bar{n}^D_1}
   ,...,
   T_{n^1_k \cdots n^D_k}, \bar{T}_{\bar{n}^1_k \cdots \bar{n}^D_k}\right)\\
   &~~~~~~=\mu \left(T_{n^1_1 \cdots n^D_1} \bar{T}_{\bar{n}^1_1 \cdots \bar{n}^D_1}
   \cdots
   T_{n^1_k \cdots n^D_k} \bar{T}_{\bar{n}^1_k \cdots \bar{n}^D_k}\right) \\
   &~~~~~~~~~~~~-\sum_{\pi} \prod_{\alpha=1}^{\alpha^{\max}}
   \kappa_{2k(\alpha)}\left(T_{n^1_{j_{\alpha}(1)} \cdots n^D_{j_{\alpha}(1)}},
   \bar{T}_{\bar{n}^1_{j_{\alpha}(1)} \cdots \bar{n}^D_{j_{\alpha}(1)}}
   ,...,
   \bar{T}_{\bar{n}^1_{j_{\alpha}(k(\alpha))} \cdots \bar{n}^D_{j_{\alpha}(k(\alpha))}}\right) \\
 \end{split}
\end{equation}
are useful. 
Here $\pi$ is a way to dividing distinct $2k$ points into 
$\alpha^{\max}$ groups ($2 \leq \alpha^{\max} \leq k$). 
As each group has $2k(\alpha)$ points ($1 \leq \alpha \leq \alpha^{\max}$), 
we see that $\sum_{\alpha=1}^{\alpha^{\max}} 2k(\alpha)=2k$. 
After expanding the average value in terms of the cumulants, 
we find the terms containing $\kappa_2$ dominate 
in the limit $N \to \infty$ and the other terms converge to zero 
with some normalization :
\begin{equation}
\begin{split}
\mu(Tr(T,\bar{T}))\sim&
\sum_{n^1,\bar{n}^1=1}^{N}
\cdots
\sum_{n^D,\bar{n}^D=1}^{N}
\left(\prod_{j=1}^k
\delta_{n^1_j \bar{n}^1_{\sigma_1 (j)}}
  \cdots
  \delta_{n^D_j \bar{n}^D_{\sigma_D (j)}}\right) \\
  &~~\times \left(\sum_{\tau \in S_k} \prod_{i=1}^k \kappa_2 
  \left(T_{n^1_i \cdots n^D_i},
   \bar{T}_{\bar{n}^1_{\tau(i)} \cdots \bar{n}^D_{\tau(i)}}\right)\right).
\end{split}
\end{equation}
Here $S_k$ is the symmetric group of degree $k$.

In order to graphically represent $\kappa_2$, 
we introduce a new edge with color $0$,  
which directly connects the vertices 
$v_i$ and $\bar{v}_{\tau(i)}$. 
Then we obtain a new $(D+1)$-colored graph $G$, 
called a covering graph of $B$ \cite{gurau1}.
Let us now assume that  
$\mu(T_{n^1 \cdots n^D} \bar{T}_{n^1 \cdots n^D})$
is one. 
Then we find
$\kappa_2(T_{n^1 \cdots n^D},\bar{T}_{\bar{n}^1 \cdots \bar{n}^D})
=\delta_{n^1 \bar {n}^1} \cdots \delta_{n^D \bar{n}^D}$, 
so that
\begin{equation}
 \begin{split}
 \mu(Tr(T,\bar{T})) \sim & \sum_{n^1,\bar{n}^1=1}^N
 \cdots
 \sum_{n^D,\bar{n}^D=1}^N 
 \left(\prod_{j=1}^k
 \delta_{n^1_j \bar{n}^1_{\sigma_1 (j)}}
  \cdots
  \delta_{n^D_j \bar{n}^D_{\sigma_D (j)}}\right) \\
  &~~\times \left(\sum_{\tau \in S_k} \prod_{i=1}^k \kappa_2 
  \left(T_{n^1_i \cdots n^D_i},
   \bar{T}_{\bar{n}^1_{\tau(i)} \cdots \bar{n}^D_{\tau(i)}}\right)\right)\\
  =&\sum_{G} N^{\sum_{i=1}^D |F^{(0,i)}(G)|}.
 \end{split}
 \end{equation}
Here $F^{(i,j)}(G)$ is the set of $(i,j)$-faces 
($2$-colored subgraphs with colors $i$ and $j$) of $G$. 
We write the cardinality of a set $S$ as $|S|$. 
When $\sum_{i=1}^D |F^{(0,i)}(G)|$ is maximum, 
$G$ is called the minimal covering graph $G^{\min}$ of $B$ \cite{gurau1}. 

Then the average trace invariants can be written 
in terms of the minimal covering graph $G^{\min}$ as
\begin{equation}
\mu(Tr(T,\bar{T})) = \sum_{G^{\min}}N^{\gamma(D)}
+o\left(N^{\gamma(D)}\right)
=|G^{\min}|N^{\gamma(D)}+o\left(N^{\gamma(D)}\right) 
\end{equation}
with
\begin{equation}
\gamma(D)=\sum_{i=1}^D |F^{(0,i)}(G^{\min})|,
\end{equation}
so that \cite{gurau1}
\begin{equation}
\lim_{N \to \infty} \frac{1}{N^{\gamma(D)}} 
\mu(Tr(T,\bar{T}))=|G^{\min}|.
\end{equation}
Here $|G^{\min}|$ is the number of the minimal covering graphs $G^{\min}$.

For general $c_1,...,c_D$, we can similarly find \cite{syuron}
\begin{equation}
\begin{split}
\mu(Tr(T,\bar{T}))\sim &
\sum_{n^1,\bar{n}^1=1}^{c_1N}
\cdots
\sum_{n^D,\bar{n}^D=1}^{c_DN} 
\left(\prod_{j=1}^k
\delta_{n^1_j \bar{n}^1_{\sigma_1 (j)}}
  \cdots
  \delta_{n^D_j \bar{n}^D_{\sigma_D (j)}}\right) \\
  &~~\times \left(\sum_{\tau \in S_k} \prod_{i=1}^k \kappa_2 
  \left(T_{n^1_i \cdots n^D_i},
   \bar{T}_{\bar{n}^1_{\tau(i)} \cdots \bar{n}^D_{\tau(i)}}\right)\right) \\
   =&\sum_{G} \prod_{i=1}^D \left(c_iN\right)^{|F^{(0,i)}(G)|} \\
\end{split}
\end{equation}
and
\begin{equation}
\begin{split}   
  \mu(Tr(T,\bar{T})) 
=&\sum_{G^{\min}} \prod_{i=1}^D \left(c_iN\right)^{|F^{(0,i)}(G^{\min})|} 
+o\left(N^{\gamma(D)}\right) \\
   =&N^{\gamma(D)}\sum_{G^{\min}}
   \prod_{i=1}^D c_i^{|F^{(0,i)}(G^{\min})|}+o\left(N^{\gamma(D)}\right) \\
\end{split}
\end{equation}
with
\begin{equation}
\gamma(D)=\sum_{i=1}^D |F^{(0,i)}(G^{\min})|,
\end{equation}
so that
\begin{equation}
\lim_{N \to \infty} \frac{1}{N^{\gamma(D)}} \mu(Tr(T,\bar{T}))
=\sum_{G^{\min}}
   \prod_{i=1}^D c_i^{|F^{(0,i)}(G^{\min})|}.
\end{equation}
Thus we have derived the asymptotic forms of the average trace invariants 
in the limit $N \to \infty$ with $c_i$ fixed. 
These forms are universal because they do not depend on the details 
of the tensor element distribution. 

In the following, we consider two special cases, 
in which one can more explicitly evaluate the asymptotic forms.
\\ \\
\noindent
{\bf 1. Melonic graphs} \\ \\
It is in general difficult to calculate $|G^{\min}|$ and 
$\sum_{i=1}^D |F^{(0,i)}(G^{\min})|$. 
However, one can explicitly evaluate them,  
when $B$ is a melonic graph defined below. 
 \begin{Def}\rm
 A dipole with $d$ colors is a $d$-colored graph 
 with only two vertices \cite{gurau1}. 
 \end{Def}
 \begin{Def}\rm(melonic graph)\\
 Suppose that $B_1$ is a $D$-dipole, 
 and that $H_i$ is the $(D-1)$-colored graph given by 
 a $D$-dipole by removing the edge with color $i$. 
 We generate a $D$-colored graph $B_k$ from $B_1$ as follows. 
  \begin{description}
  \item[\rm (1)]To obtain $B_2$, cut the edge with color $i_1$ on $B_1$
  and insert $H_{i_1}$ there.
  \item[\rm (2)]To obtain $B_3$, cut one of the two edges with color $i_2$ 
  on $B_2$ and insert $H_{i_2}$ there.
  \item[\rm (3)] To obtain $B_4$, cut one of the edges with color $i_3$ 
  on $B_3$ and insert $H_{i_3}$ there.\\
  \vdots
  \item[\rm (k-1)] To obtain $B_k$, cut one of the edges with color $i_{k-1}$ 
  on $B_{k-1}$ and insert $H_{i_{k-1}}$ there.
  \end{description}
  A $D$-colored graph $B_k$ (with $2k$ vertices) 
  given above is called a melonic graph \cite{gurau1}.
 \end{Def}
 \begin{Lem}\rm \cite{gurau1}
 A melonic $D$-colored graph has only one minimal 
 covering graph $G^{\min}$ satisfying 
 $\sum_{i=1}^D |F^{(0,i)}(G^{\min})|=1+k(D-1)$.
 \end{Lem}
 Therefore the universal asymptotic forms of the average 
 trace invariants corresponding to melonic graphs can be 
 calculated as
 \begin{equation}
 \lim_{N \to \infty} \frac{1}{N^{1+k(D-1)}}\mu(Tr(T,\bar{T}))
 =\prod_{i=1}^D c_i^{|F^{(0,i)}(G^{\min})|}.
 \end{equation}
 However, in order to have explicit asymptotic forms, 
 we still need to know the value of each $|F^{(0,i)}(G^{\min})|$, 
 except for the case $c_1 = \cdots = c_D$.
 \\
  \\
   \noindent
 {\bf 2. Cycle graphs with multiple edges} \\ \\
 When $R$ is a cycle graph with multiple edges, 
 we are able to explicitly evaluate 
 the asymptotic universal forms of the average trace invariants 
 for general $c_1,...,c_D$. 
 Let us begin with the definition of such a cycle graph.
 \begin{Def}\rm
 Let $m,n \in \mathbb{N}$ satisfy $m+n=D$.
 An $(m,n)$-cycle graph $R$ is 
 defined as a $D$-colored graph with $2k$ 
 vertices satisfying the following conditions.
  \begin{description} 
  \item[\rm (1)] $k$ $m$-dipoles and $k$ $n$-dipoles are the subgraphs of $R$. 
  \item[\rm (2)]$m$-dipoles have $m$ edges with colors $i_1,...,i_m$ 
  and $n$-dipoles have $n$ edges with colors $i'_1,...,i'_n$, 
  provided that $i_{\nu} \neq i'_{\nu'}$ for all $\nu,\nu'$. 
  \end{description}
 \end{Def}
\noindent
A $D$-colored graph with $2k$ vertices has $Dk$ edges. 
Thus $R$ does not have any edges except the dipole edges. 
Therefore it is a cycle graph with 
multiple edges composed of $2k$ dipoles
(Figure \ref{2,3moko}).
 \begin{figure}[h!!]
  \begin{center}
  \includegraphics[width=4cm]{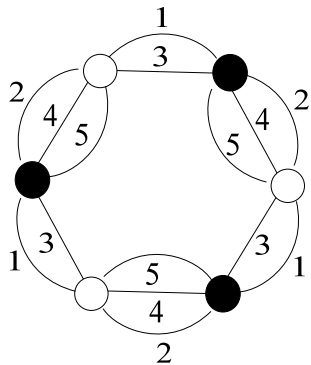}
  \caption{A $(2,3)$-cycle graph $R$ with $6$ vertices.}
  \label{2,3moko}
  \end{center}
 \end{figure}

Note that $(1,1)$-cycle graphs represent trace 
invariants of random matrices, and that $(1,n)$-cycle 
graphs are melonic when $n \neq 1$. 
We can count the number of the minimal covering graphs 
and their $(0,j)$-faces for such $(1,n)$-cycle graphs. 
This fact helps us understand $(m,n)$-cycle graphs for general $m$ and $n$. 

\begin{figure}[h!!]
   \begin{center}
   \includegraphics[width=8cm]{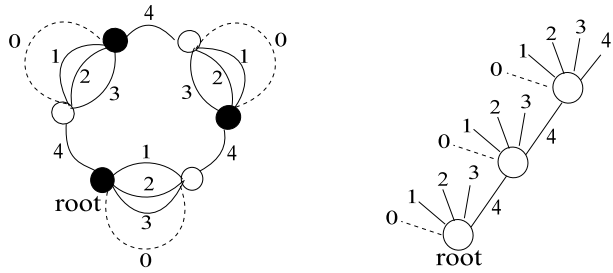}
   \caption{The only minimal covering graph of $R$.}
   \label{minimal}
   \end{center}
  \end{figure}

\begin{figure}[h!!]
  \begin{center}
  \includegraphics[width=12cm]{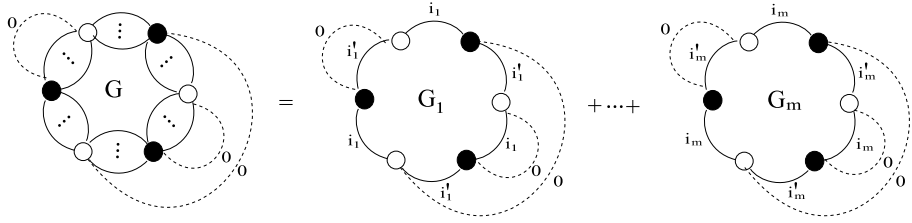}
  \caption{A covering graph $G$ of $R$.}
  \label{Gdivided}
  \end{center}
 \end{figure}

 \begin{Lem}\rm
 Suppose that a $(1,1)$-cycle graph $R$ has $2k$ vertices.
 \begin{description}
 \item[\rm (1)]
 The number of the minimal covering graphs of $R$ is the Catalan number\cite{catalan} 
 $C_k = \frac{1}{k+1} \binom{2k}{k}$.
 The total number of $(0,i)$-faces 
 ($1 \leq i \leq D$) of each minimal covering graph is $k+1$.
\item[\rm (2)] The number of the minimal covering graphs (of $R$) which have $l$ $(0,1)$-faces 
 is the Narayana number\cite{narayana} $N_{k,l}=\frac{1}{k} \binom{k}{l} \binom{k}{l-1}$.
 \end{description}
 \end{Lem}
 \noindent
 Proof. 
 \begin{description}
 \item[\rm (1)]
 Let $G$ be a covering graph of $R$ with $2k$ vertices and $3k$ edges. 
 Then
  \begin{equation}
  \label{euler}
   \begin{split}
   &\sum_{0 \leq i < j} |F^{(i,j)}(G)|-|E(G)|+|V(G)| \\
   =&\sum_{i=1}^2 |F^{(0,i)}(G)|+1-k \\
   =&2-2g(G). \\
   \end{split}
  \end{equation}
 Thus if there is a covering graph $G$ satisfying $g(G)=0$ (i.e. $G$ are planar), 
 it is a minimal covering graph $G^{\min}$.
 The edges with color $0$ of a planar $G$ give a plane partition of $R$.
 The number of the plane partitions of $R$, that is $|G^{\min}|$, 
 is the Catalan number $C_k$.
 Moreover it follows from eq. (\ref{euler}) that
 \begin{equation}
 \sum_{i=1}^2 |F^{(0,i)}(G^{\min})|=k+1.
 \end{equation}
 \\ \\
  \begin{minipage}{0.5\textwidth}
    \begin{center}
    \includegraphics[width=3cm]{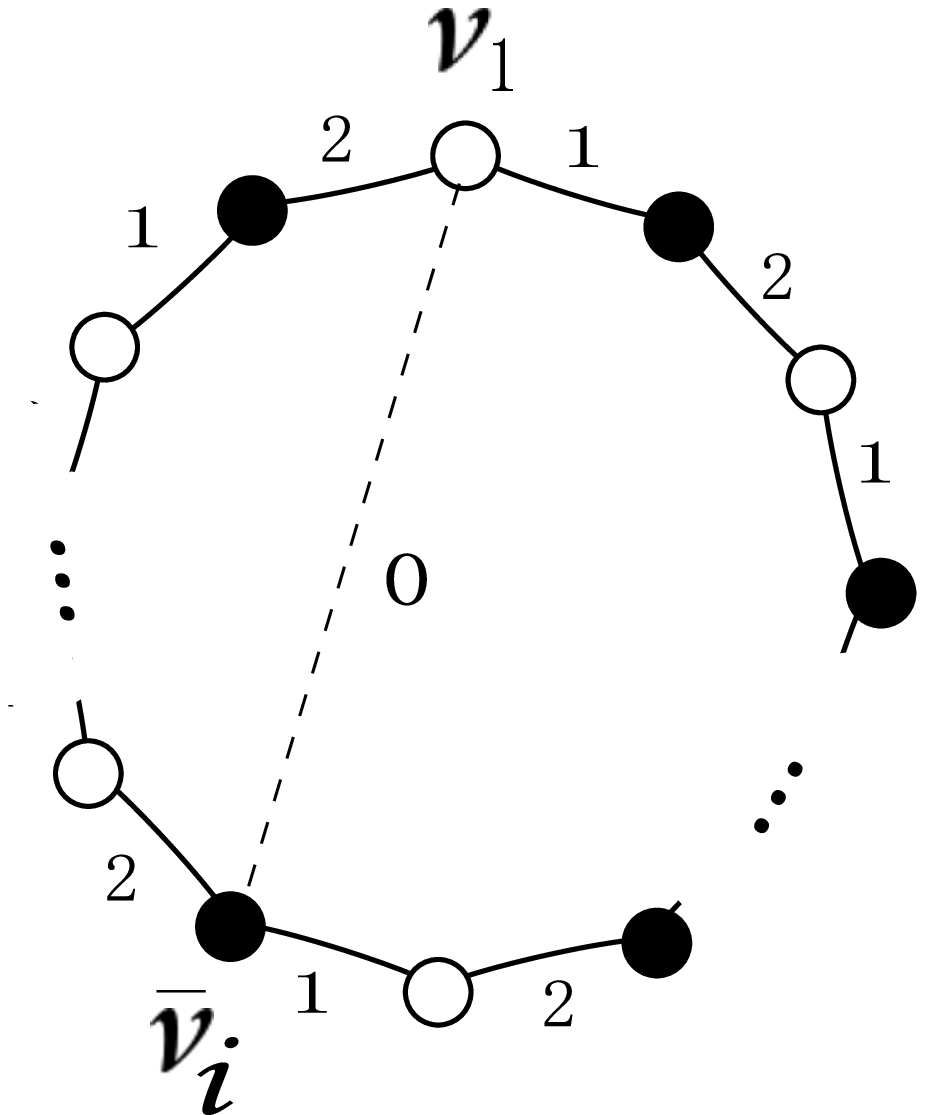}
    \label{catalan}
    \end{center}
  \end{minipage}
 \hspace*{-0.5cm}
  \begin{minipage}[t]{0.5\textwidth}
  Dividing a cycle graph at an edge $(v_1,{\bar v}_i)$ with color $0$ gives 
  a recurrence formula for $C_k$. 
  Here $v_1$ is fixed and $i \in \{ 1,...,k \}$.
  \end{minipage}
 \item[\rm (2)]
 Let $N_{k,l}$ be the number of minimal covering graphs having $l$ $(0,1)$-faces.
 We fix an edge $e$ (on the $(1,1)$-cycle graph $R$) 
 with color $1$ and denote the $(0,1)$-face containing
 the edge $e$ by $F$.
 Suppose that $F$ has $2m$ vertices ($m \in \{ 1,...,k \}$).
 Then $R$ is divided into $(m+1)$ parts 
 ($B_1,...,B_m$ and $F$).
 The number of $B_i$ having $2k_i$ vertices and $l_i$ $(0,1)$-faces
 is $N_{k_i,l_i}$.
 Thus we obtain a recurrence formula\cite{syuron,narayana}
 \begin{equation}
 \begin{split}
  N_{k,l}&=\sum_{k_1=k-1} ~\sum_{l_1=l-1}N_{k_1,l_1}
  +\sum_{k_1+k_2=k-2} ~\sum_{l_1+l_2=l-1}N_{k_1,l_1}N_{k_2,l_2} \\
  &~~~~~~~~~~~~~~~~~~~~~
  +\sum_{k_1+k_2+k_3=k-3} ~\sum_{l_1+l_2+l_3=l-1}N_{k_1,l_1}N_{k_2,l_2}N_{k_3,l_3}
  + \cdots.
  \end{split}
 \end{equation}
 We can make use of the Lagrange inversion formula\cite{hantenkousiki}
 to derive $N_{k,l}=\frac{1}{k} \binom{k}{l} \binom{k}{l-1}$
 from this recurrence formula.
 \qed 
 \end{description}
 \begin{Lem}\rm
 \label{lem1,nmoko}
 When $n \neq 1$, a $(1,n)$-cycle graph $R$ with $2k$ vertices 
 has only one minimal covering graph.
 The total number of $(0,i)$-faces 
 $(1 \leq i \leq D)$ of the minimal covering graph is $nk+1$.
 \end{Lem}
 \noindent
 Proof.
 We illustrate in Figure \ref{1,nmoko} that $R$ is melonic.
 Its only minimal covering graph $G^{\min}$ is depicted in Figure \ref{minimal}.
  \begin{figure}[h!!]
   \begin{center}
   \includegraphics[width=12cm]{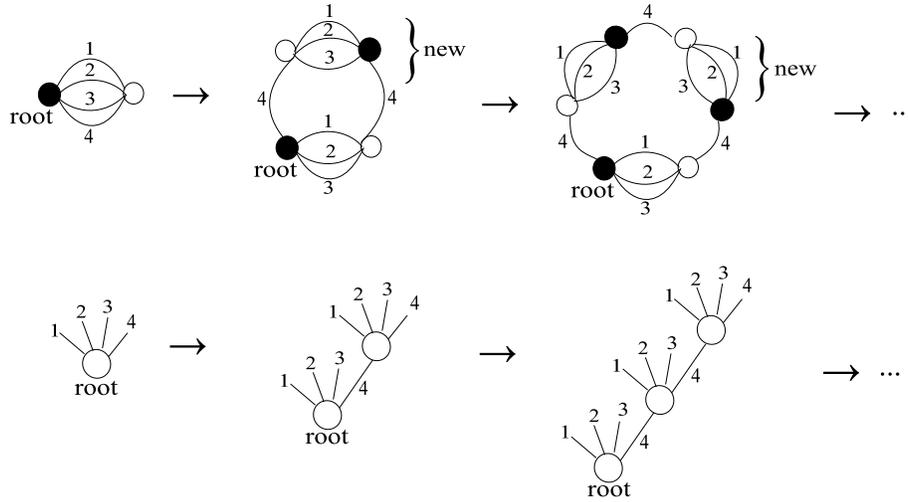}
   \caption{A $(1,n)$-cycle graph constructed as a melonic graph.}
    \label{1,nmoko}
   \end{center}
  \end{figure}
  \noindent
  We see that $G^{\min}$ is a $(1,n+1)$-cycle graph,
  and easily find 
 \begin{equation}
 \sum_{i=1}^D |F^{(0,i)}(G^{\min})|=nk+1.
 \end{equation}
 \qed
\\ \\
We are now in a position to consider a general $(m,n)$-cycle graph $R$. 
Two cases (the case $m=n$ and the case $m<n$) are separately treated. 
We first evaluate $|G^{\min}|$ and $|F^{(0,i)}(G^{\min})|$,
and then explicitly calculate the average trace invariants. 
 \\ \\
\noindent
$\bullet$ The case $m=n$.
An $(m,n)$-cycle graph $R$ has $2k$ $m$-dipoles, 
$k$ of which have edges with colors $i_1,...,i_m$ 
and the others $i'_1,...,i'_m$.
Since the dipoles with colors $i_1,...,i_m$ and those with $i'_1,...,i'_m$ 
are alternately connected in $R$,
$R$ is divided into $m$ $(1,1)$-cycle graphs 
with colors $i_{\nu}$ and $i'_{\nu}$ ($\nu=1,...,m$).
\begin{figure}[h!!]
  \begin{center}
  \includegraphics[width=12cm]{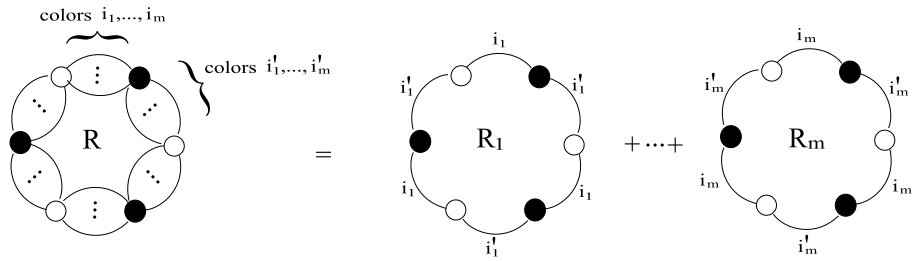}
  \caption{An $(m,n)$-cycle graph $R$ made of $m$ cycles.}
  \label{Rdivided}
  \end{center}
 \end{figure}
\\See Figure \ref{Rdivided}.
Let us define $R_{\nu}$ as the $(1,1)$-cycle graph 
with colors $i_{\nu}$ and $i'_{\nu}$. 
The covering graph $G$ of $R$ is also divided into graphs $G_{\nu}$ $(\nu = 1,...,m)$. 
Here $G_{\nu}$ is the covering graph of $R_{\nu}$. 
See Figure \ref{Gdivided}. 

The number of $(0,i)$-faces of $G$ is equal to the total number of 
$(0,i)$-faces of $G_{\nu}$ :
 \begin{equation}
 \sum_{i=1}^D |F^{(0,i)}(G)|
 =\sum_{\nu=1}^m \sum_{i=i_{\nu},i'_{\nu}} |F^{(0,i)}(G_{\nu})|.
 \end{equation}
Since all $G_{\nu}$ have the same form (except their colors) 
for each covering graph $G$, 
if $G$ is the minimal covering graph of $R$,
$G_{\nu}$ is the minimal covering graph of $R_{\nu}$ for each $\nu$. 
Therefore
\begin{equation}
 |G^{\min}|=|G^{\min}_{1}|=\cdots=|G^{\min}_{m}|=C_k.
\end{equation}
Here $G^{\min}_{\nu}$ is the minimal covering graph of $R_{\nu}$. 
One can also see that
\begin{equation}
 \begin{split}
 \sum_{i=1}^D |F^{(0,i)}(G^{\min})|
 =&\sum_{\nu=1}^m \sum_{i=i_{\nu},i'_{\nu}}
 |F^{(0,i)}(G_{\nu}^{\min})| \\
 =&\sum_{\nu=1}^m(k+1) \\
 =&m(k+1). \\
 \end{split}
\end{equation}

Now the universal asymptotic formula for the average trace invariants $\mu(Tr(T,{\bar{T}}))$ 
of $N \times \cdots \times N$ random tensors $T$ is written in the form 
\begin{equation}
 \begin{split}
 &\frac{1}{N^{\gamma(D)}}
 \mu(Tr(T,\bar{T})) \\
 =&\frac{1}{N^{\gamma(D)}}
 \left(|G^{\min}|N^{\gamma(D)}
 +o\left(N^{\gamma(D)}\right)\right) \\
 =&\frac{1}{N^{m(k+1)}} \left(C_k N^{m(k+1)} 
 +o\left(N^{m(k+1)}\right)\right)\\
 &\longrightarrow ~~C_k ~~~~~(N \to \infty).
 \end{split}
\end{equation}
The more general formula for $c_1N \times \cdots \times c_DN$ tensors $T$ is
\begin{equation}
\label{narayana}
 \begin{split}
 &\frac{1}{N^{\gamma(D)}}
 \mu(Tr(T,\bar{T})) \\
 =&\frac{1}{N^{\gamma(D)}}
 \left(\sum_{G^{\min}} \prod_{j=1}^D (c_jN)^{|F^{(0,j)}(G^{\min})|} 
 +o\left(N^{\gamma(D)}\right)\right)\\
 =&\frac{1}{N^{m(k+1)}}
 \left(\sum_{l=1}^k N_{k,l}
 \prod_{\nu=1}^m c_{i_{\nu}}^l c_{i'_{\nu}}^{k-l+1}
 N^{m(k+1)} 
 +o\left(N^{m(k+1)}\right)\right)\\
 &\longrightarrow ~~\sum_{l=1}^k N_{k,l}
 \prod_{\nu=1}^m c_{i_{\nu}}^l  c_{i'_{\nu}}^{k-l+1}
 ~~~~~(N \to \infty).
 \end{split}
\end{equation}
Here the minimal covering graphs $G^{\min}_{\nu}$ 
are classified according to the number 
$l=|F^{(0,i_{\nu})}(G^{\min}_{\nu})|$. 
Then the number of $G^{\min}_{\nu}$ 
is given by the Narayana number $N_{k,l}$. 
We see that eq. (\ref{narayana}) follows from the equality
$|F^{(0,i_1)}(G_1^{\min})|= \cdots =|F^{(0,i_m)}(G_m^{\min})|$.
 \\ \\
 \noindent
 $\bullet$ The case $m<n$.
An $(m,n)$-cycle graph $R$ has $k$ $m-$dipoles with colors $i_1,...,i_m$ and 
$k$ $n$-dipoles with colors $i'_1,...,i'_n$. 
As before $R$ can be divided into $(m-1)$ $(1,1)$-cycle graphs $R_{\nu}$ 
with colors $i_{\nu}$ and $i'_{\nu}$ ($\nu \in \{ 1,...,m-1 \}$). 
and a $(1,n-m+1)$-cycle graph $R_m$. 
See Figure \ref{n,mmoko}. 
 \begin{figure}[h!!]
  \begin{center}
  \includegraphics[width=12cm]{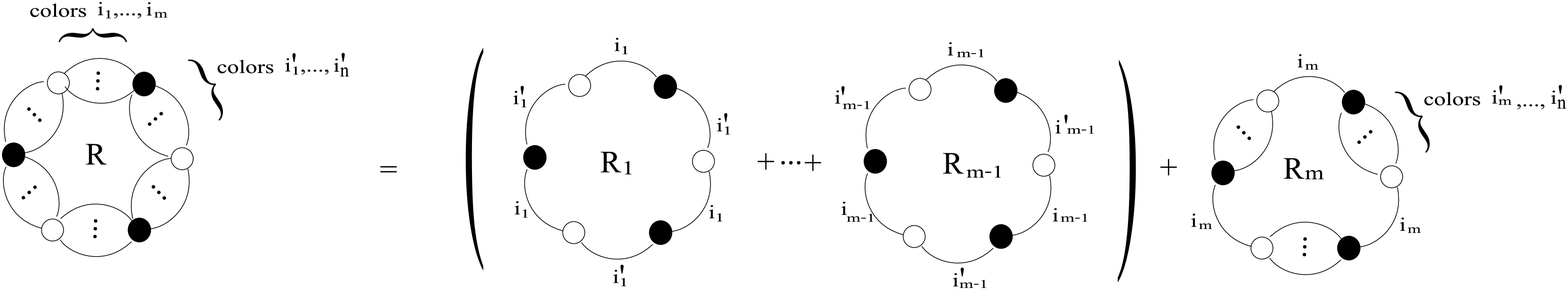}
  \caption{}
  \label{n,mmoko}
  \end{center}
 \end{figure}
 The covering graphs $G$ are also divided into $m$ graphs $G_{\nu}$.
  See Figure \ref{n,mmoko2}.
 \begin{figure}[h!!]
  \begin{center}
  \includegraphics[width=12cm]{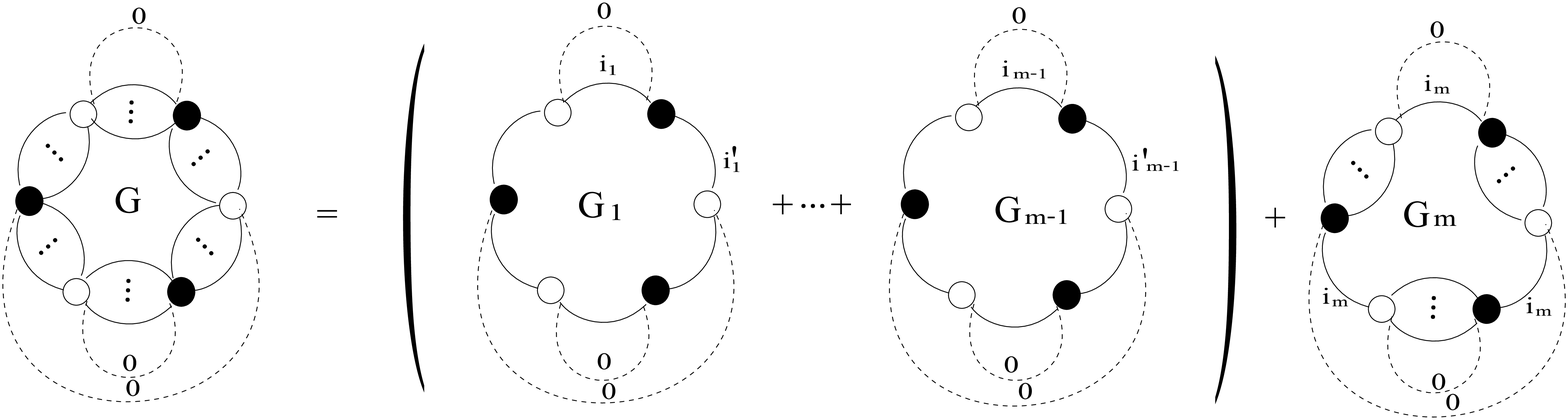}
  \caption{}
  \label{n,mmoko2}
  \end{center}
 \end{figure}
 All $G_{\nu}$ can be the minimal covering graphs of $R_{\nu}$ at the same time, 
 because, if $G_m$ is minimal, all $G_{\nu}$ for $\nu \in \{ 1,...,m-1 \}$ are minimal 
(Check the form of the minimal covering graph of a $(1,n)$-cycle graph in Lemma \ref{lem1,nmoko}).
Since $R_m$ has only one minimal covering graph (Lemma \ref{lem1,nmoko}), we find
\begin{equation}
|G^{\min}|=1.
\end{equation}
The number of $(0,i)$-faces of $G^{\min}$ is
\begin{equation}
\begin{split}\sum_{i=1}^D |F^{(0,i)}(G^{\min})|
&=\sum_{\nu=1}^m \sum_{i=1}^D |F^{(0,i)}(G^{\min}_{\nu})| \\
&=\sum_{\nu=1}^{m-1} (k+1) +(n-m+1)k+1 \\
&=nk+m.\\
\end{split}
\end{equation}
Here $G^{\min}_{\nu}$ is the minimal covering graph of $R_{\nu}$ ($\nu \in \{ 1,...,m \}$).
Then the universality for $N \times \cdots \times N$ tensors $T$ is 
given by the formula
\begin{equation}
 \begin{split}
 &\frac{1}{N^{\gamma(D)}}
 \mu(Tr(T,\bar{T})) \\
 =&\left(\frac{1}{N^{\gamma(D)}}
 |G^{\min}|
 N^{\gamma(D)} 
 +o\left(N^{\gamma(D)}\right)\right)\\
 =&\frac{1}{N^{nk+m}}
 \left(N^{nk+m}
 +o\left(N^{nk+m}\right)\right)\\
 &\longrightarrow ~~1~~~~~(N \to \infty).
 \end{split}
\end{equation}
More generally, for $c_1N \times \cdots \times c_DN$ tensors $T$, 
we obtain
\begin{equation}
\label{narayana2}
 \begin{split}
 &\frac{1}{N^{\gamma(D)}}
 \mu(Tr(T,\bar{T})) \\
 =&\frac{1}{N^{\gamma(D)}}
 \left(\sum_{G^{\min}} \prod_{j=1}^D \left(c_jN\right)^{|F^{(0,j)}(G^{\min})|}
 +o\left(N^{\gamma(D)}\right)\right)\\
 =&\frac{1}{N^{nk+m}}
 \left(N^{nk+m}
 \left(\prod_{\nu=1}^m c_{i_{\nu}}\right) \left(\prod_{\nu'=1}^n c_{i'_{\nu'}}^k\right)
 +o\left(N^{nk+m}\right)\right)\\
 &\longrightarrow ~~\left(\prod_{\nu=1}^m c_{i_{\nu}}\right)
 \left(\prod_{\nu'=1}^n  c_{i'_{\nu'}}^k\right)
 ~~~~~(N \to \infty).
 \end{split}
\end{equation}

In summary, we studied the universal asymptotic forms in the limit
$N \to \infty$ of the average trace invariants of $c_1N \times \cdots \times c_DN$
random tensors.
In the special cases corresponding to cycle graphs with multiple edges, 
we explicitly calculated the asymptotic forms.

\section*{Acknowledgement}
The author thanks Taro Nagao for valuable suggestions and discussions.

\end{document}